\documentclass[aps,prb,twocolumn,groupedaddress,showpacs]{revtex4-1}
\usepackage{graphicx}
\begin{document}

\title{Superconducting Properties of the K$_{{x}}$WO$_{3}$ Tetragonal Tungsten Bronze and the Superconducting Phase Diagram of the Tungsten Bronze Family}

\author{Neel Haldolaarachchige, Quinn Gibson, Jason Krizan and R.~J.~Cava}
\affiliation{Department of Chemistry, Princeton University, Princeton, New Jersey 08544, USA}

\begin{abstract}  
We report the superconducting properties of the K$_{x}$WO$_{3}$ tetragonal tungsten bronze. The highest superconducting transition temperature ($T_{c}=2.1$~K) was obtained for K$_{0.38}$WO$_{3}$. $T_{c}$ decreases linearly with increasing K content. Using the measured values for the upper critical field $H_{c2}$, and the specific heat $C$, we estimate the orbital critical field $H_{c2}$(0), coherence length $\xi$(0), Debye temperature $\Theta  _{D}$ and coupling constant $\lambda  _{ep}$. The magnitude of the specific heat jump at $T_{c}$ suggests that the K$_{x}$WO$_{3}$ tetragonal tungsten bronze is a weakly-coupled superconductor. The superconducting phase diagram of the doped tungsten bronze family is presented.  
  
\end{abstract}


\maketitle
\newcommand{\angstrom}{\mbox{\normalfont\AA}}

\section{Introduction}
The $M_{x}$WO$_{3}$ tungsten bronzes, where $M$ is an ion that donates electrons to the WO$_{3}$ framework, show a wide range of interesting physical properties and several distinct crystal structures. Many studies have been performed on the superconductivity in this family since its initial discovery.~\cite{mattias1964} Tungsten bronzes share features with other oxide superconducting systems, such as having a relatively low density of electronic states at the Fermi Energy and $T_{c}$s that are highest close to a structural phase boundary. The superconducting $M_{x}$WO$_{3}$ tungsten bronzes are non-stoichiometric compounds that can have hexagonal bronze ($i.e.$~HTB), tetragonal bronze (TTB-I), or perovskite-derived tetragonal (TTB-II) and cubic (CTB) structures.~\cite{sweedler1965,2sweedler1965,mattias1964,sleight1975,brusetti2007,cadwell1981,wiseman1973,ostenson1978,kasl2013,lee1997,sagar2010,ingham2005,
2ting2007,skakon1979,stanley1979, howard1975, guo2008,banks1968,remeika1967, daigo2011,josh2013,shein2011}

Although the superconducting HTBs in particular have been well studied, a detailed investigation of the superconducting properties of the tetragonal tungsten bronze (TTB-I) system is lacking up to the present time. Superconductivity in TTB-I bronzes has been reported for Na$_{x}$WO$_{3}$ $0.2 < x (Na) < 0.5$, K$_{x}$WO$_{3}$ $0.4 < x (K) < 0.57$ and Ba$_{0.13}$WO$_{3}$,~\cite{howard1975,sweedler1965,2sweedler1965} 
but contradicting this, TTB-I K$_{0.58}$WO$_{3}$ has recently been reported as a semiconductor.~\cite{fan2000} Here we report a detailed study on the superconducting K$_{x}$WO$_{3}$ tetragonal tungsten bronze TTB-I phase. The experimental characterization is supplemented by electronic structure calculations. Finally, we present a phase diagram for superconductivity across the full HTB, TTB-I, and CTB tungsten bronze family.

\section{Experiment and Calculation}
The K$_{x}$WO$_{3}$ samples were prepared by solid state reaction method starting from pre-made K$_{2}$WO$_{4}$, purified WO$_{2}$ (99.8$\%$; Alfa Aesar) and WO$_{3}$ (99.8$\%$; Alfa Aesar) according to the following reaction:

\begin{equation}
\label{equ1} \frac{x}{2}K_{2}WO_{4}+(1-x)WO_{3}+\frac{x}{2}WO_{2} \rightarrow K_{x}WO_{3}
\end{equation}

The K$_{2}$WO$_{4}$ precursor was prepared by reacting stoichiometric amounts of K$_{2}$CO$_{3}$ and WO$_{3}$ in air at 650~$^{0}$C for 12 hrs. Stoichiometric amounts of materials according to equation 1 were mixed and pressed into pellets inside an Ar-filled glove box. The pellets were transferred into alumina crucibles and sealed inside evacuated quartz glass tubes that were heated at 650~$^{0}$C and 800~$^{0}$C for two nights with intermediate grinding. The reacted samples were kept inside the glove box. (Such handling is necessary to avoid the decomposition that possibly results in the observation of semiconducting behavior.) The purity and cell parameters of the samples were evaluated by powder X-ray diffraction (PXRD) data at room temperature on a Bruker D8 FOCUS diffractometer (Cu$~K_{\alpha})$. Lattice parameters were refined by the Rietveld method~\cite{rietveld1969} using the FULLPROF program integrated in the WINPLOTR software.~\cite{fullprof} Elemental analysis performed using energy dispersive spectroscopy (EDS) showed that the potassium contents in the samples are very close to the nominal values; therefore the nominal potassium stoichiometries are used in this report. 

The electrical resistivity was measured using a standard four-probe dc-technique with an excitation current of 10 mA; small diameter Pt wires were attached to the sample using silver paste. Data were collected from 0.4 K to 300 K in zero-field and also in magnetic fields up to 1 T using a Quantum Design Physical Property Measurement System (QD-PPMS) equipped with a $^{3}$He cryostat. The specific heat was measured between 0.4 K and 20 K under fields of $\mu _0$H = 0 T and 1 T in the PPMS using the time-relaxation method.
The electronic structure calculations were performed by density functional theory (DFT) using the WIEN2K code with a full-potential linearized augmented plane-wave and local orbitals [FP-LAPW + lo] basis~\cite{blaha2001,sign1996,madsen2001,sjosted2000} together with the PBE parameterization~\cite{perdew1996} of the GGA, including spin orbit coupling (SOC). The plane-wave cutoff parameter R$_{MT}$K$_{MAX}$ was set to 7 and the Brillouin zone was sampled by 2000 k points. Experimental lattice parameters from the Rietveld refinements were used in the calculations. A fully ordered K$_{0.6}$WO$_{3}$ TTB crystal structure (K$_6$W$_{10}$O$_{30}$) was employed for the calculations. 

\section{Results and Discussion}
\begin{figure}[t]
  \centerline{\includegraphics[width=0.5\textwidth]{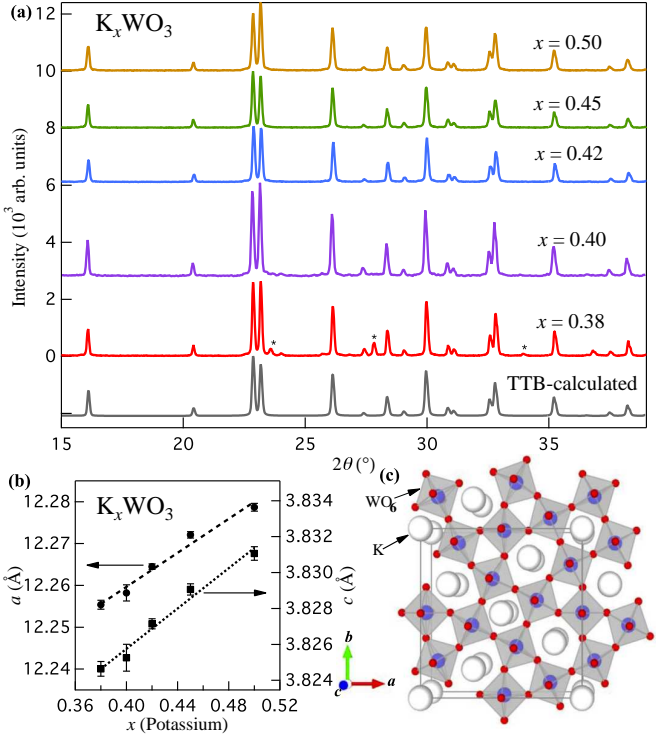}}
  \caption
    {
      (Color online) Analysis of the X-ray diffraction data for the K-doped TTB-I bronze. (a) PXRD patterns of the K$_{x}$WO$_{3}$ samples in the range of $0.38 \le x \le 0.50$. Asterisk marks in $x$ = 0.38 PXRD pattern indicates impurity peaks from HTB phase. (b) Lattice parameters $a$ and $c$ as a function of K doping. (c) The crystal structure of the tetragonal tungsten bronze superconductor.
    }
  \label{Fig1}
\end{figure}

\begin{figure}[t]
  \centerline{\includegraphics[width=0.5\textwidth]{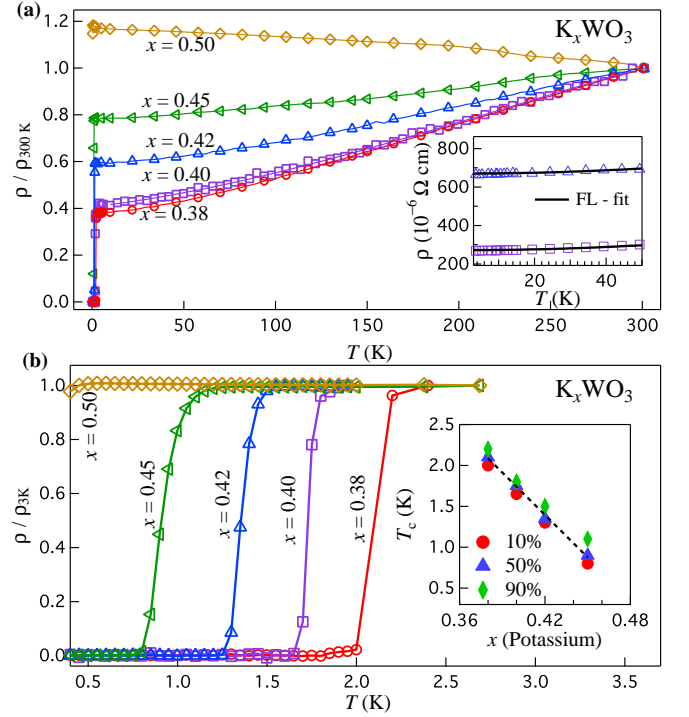}}
  \caption
    {
      (Color online) Analysis of temperature dependent resistivity data for the K doped TTB-I bronze. (a) Temperature dependent normalized resistivity. Inset shows temperature dependent resistivity for the K$_{x}$WO$_{3}$, $x$ = 0.40 and 0.42 samples. The solid line represents the Fermi-liquid fit, with $\rho =\rho _{0}+AT^{n}$. (b) Low temperature superconducting transition for $0.38 \le x \le 0.50$. Inset shows the normal-state resistance drop of 10$\%$, 50$\%$ and 90$\%$ as a function of doping. The dashed line represents linear fit to the 50$\%$ drop point.   
    }
  \label{Fig2}
\end{figure}

Fig.~\ref{Fig1} shows the structural analysis of the materials by PXRD. Samples from $x~=~0.40$ to $x~=~0.50$ are purely the tetragonal-I phase (P4/$mbm$, space group number 127) without any impurities. The $x=0.38$ sample shows a small amount of HTB impurity (asterisk marks in Fig.~\ref{Fig1}(a)). A linear variation of lattice parameters with composition can be observed in Fig.~\ref{Fig1}(b). This behavior is a good indicator that the TTB-I series is successfully formed in this composition range. The $x=0.38$ sample falls very well on the trend, and also on the trend of the linear variation of $T_{c}$ with potassium doping (see inset of Fig.~\ref{Fig2}(b)). These observations, in addition to the presence of the small amount of impurity phase in the PXRD pattern indicates that $x=0.38$ is very close to the HTB-TTB phase boundary. The crystal structure of the K$_{x}$WO$_{3}$ tetragonal tungsten bronze is shown in Fig.~\ref{Fig1}(c).

WO$_3$ itself is a semiconductor with a band gap of approximately 2.5 eV.~\cite{kopp1977} For the tungsten bronzes, alkali metals such as K or Na in the tunnels donate one electron per ion to the WO$_3$ framework, inducing metallic behavior and superconductivity in the framework at a sufficient electron doping level. Fig.~\ref{Fig2} shows the electrical resistivity data for the  K$_{x}$WO$_{3}$ TTB-I phase in the range of $0.38 \le x \le 0.50$. Fig.~\ref{Fig2}(a) shows the normalized resistivity from 300 K to 0.4 K. The normal state resistance behavior of the TTB-I system is quite different from that of the HTB system, which shows anomalous peaks in $R$(T) and semiconducting behavior in some composition ranges.~\cite{cadwell1981,josh2013} The present system is a bad metal, showing an almost temperature independent resistance in the normal state for most compositions of the TTB-I phase, with a residual resistivity ratio ($RRR=\frac{R_{300 K}}{R_{2 K}}$) that decreases with increasing K doping. This may indicate that the disordered K in the tunnels (which would be filled for K$_{0.60}$WO$_{3}$) has a stronger influence on the resistivity at higher electron doping levels. 

The temperature dependence of the resistivites for the $x=0.40$ and $x=0.42$ samples between 2 K and 50 K are shown in the inset for Fig.~\ref{Fig2}(a). The behavior is metallic $\left( \frac{d\rho}{dT} > 0\right)$. The normal-state resistivity is fairly low (about $1~m\Omega \cdot cm$ at room temperature) and the residual resistivity ratio ($RRR = 1.7$) is small. The low-temperature resistivity data can be described by the power law, $\rho = \rho _{0} + AT^{n}$ with $n=2$, the residual resistivity $\rho _{0} = 0.27~m \Omega ~cm$, and the coefficient $A = 0.00001~\mu \Omega ~cm/K^{2}$. The Fermi-liquid fits are shown as
solid lines in the inset to Fig.~\ref{Fig2}(a). The value of $\rho _{0}$ is small, an indication of the good quality of the polycrystalline samples employed for the measurements.
The value of $A$ is often taken as a measure of the degree of electron correlations; the value found here suggests that K$_{x}$WO$_{3}$ is a weakly correlated electron system and the variation of the low temperature resistivity with $T^{2}$ indicates Fermi-liquid behavior.~\cite{kittel}
Fig.~\ref{Fig2}(b) shows the superconducting transition for all the K-doped TTB-I bronze samples. The quality of the superconducting transition is high - the 10~$\%$ to 90~$\%$ width for all the samples is smaller than 0.2 K. The raw data clearly show that the superconducting $T_{c}$ decreases with increasing electron doping (i.e. increasing K content) of the TTB-I phase. The behavior of $T_{c}$ with doping is extracted from these data and is presented in the inset of Fig.~\ref{Fig2}(b).

\begin{figure}[t]
  \centerline{\includegraphics[width=0.5\textwidth]{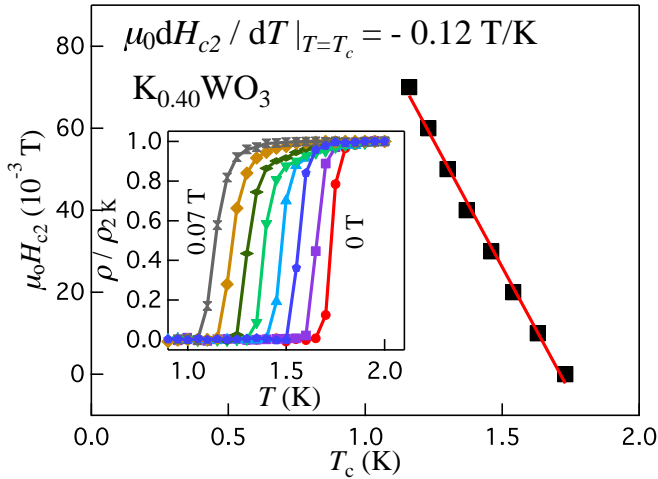}}
  \caption
    {
      (Color online) Magnetoresistance analysis of K$_{0.40}$WO$_{3}$ sample. Main panel shows $\mu _{0}H_{c2}$ as a function of $T_{c}$ and the inset shows resistivity as a function of temperature with applied magnetic field. The solid line represents liner fit. 
      %WHH-fit with $\mu _{0}H_{c2}=\mu _{0}H_{c2}(0)\left[ 1-\left( \frac{T}{T_{c}(H=0)}\right)^{\frac{3}{2}}\right] ^{\frac{3}{2}}$.   
    }
  \label{Fig3}
\end{figure}

Fig.~\ref{Fig3} shows an analysis of the magnetoresistance data for the $x=0.40$ sample. The width of the superconducting transition increases slightly with increasing magnetic field. Selecting the 50$\%$ normal resistivity point as the transition temperature, we estimate the orbital upper critical field, $\mu _{0}H_{c2}$(0), from the Werthamer-Helfand-Hohenberg (WHH) expression, 
$\mu _{0}H_{c2}(0)=-0.693~T_{c}\frac{dH_{c2}}{dT}\vert_{T=T_{c}}$. A very linear relationship is observed in the Fig.~\ref{Fig3} between $\mu _{0}H_{c2}$ and $T_{c}$. The slope is used to calculate $\mu _{0}H_{c2}(0)=$~0.14~T for $x=0.40$. The value of $\mu _{0}H_{c2}(0)$ is much smaller than the weak coupling Pauli paramagnetic limit $\mu _{0}H^{Pauli} = 1.84~T_{c} = 3.22$ T for $x=0.40$. Small values of upper critical field are an intrinsic property of the superconducting tungsten bronze family; the upper critical field ($\mu _{0}H_{c2}(0)$) that we find for the K$_{x}$WO$_{3}$ TTB-I superconductor is lower than that reported for the HTB and CTB systems (see Table.~\ref{tab:2}), with the exception of the Rb doped HTB,~\cite{ting2007} which shows similar values of $H_{c2}(0)$. The upper critical field value $\mu _{0}H_{c2}(0)$ of the K$_{0.40}$WO$_{3}$ TTB-I superconductor can be used to estimate the Ginzburg-Landau coherence length $\xi (0)=\sqrt{\Phi _{0}/2\pi H_{c2}(0)}=485$~\AA, where $\Phi _{0}=\frac{hc}{2e}$ is the magnetic flux quantum.~\cite{clogston1962,werthamer1966} This value is comparable to that for the Rb doped HTB and larger than that for the F doped CTB.~\cite{daigo2011,ting2007}

\begin{figure}[t]
  \centerline{\includegraphics[width=0.5\textwidth]{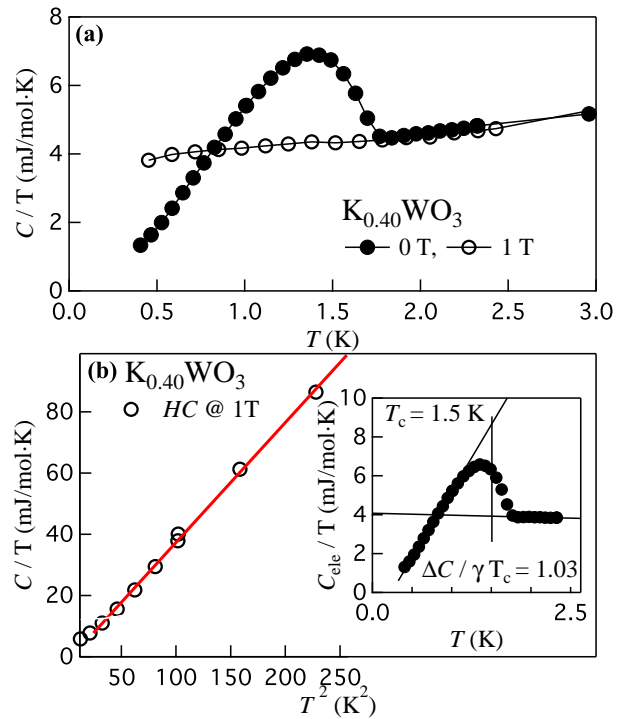}}
  \caption
    {
      (Color online) Analysis of heat capacity of K$_{0.40}$WO$_{3}$ sample. (a) Temperature dependent heat capacity at zero field and at 1 T field. (b) $\frac{C}{T}$ as a function of $T^{2}$ and solid line represent the fit with $\frac{C}{T}=\gamma + \beta T^{2}$. Inset of (b) shows equal area construction to find $T_{c}$ and $\frac{\Delta C}{\gamma T}$.   
    }
  \label{Fig4}
\end{figure}

Fig.~\ref{Fig4} shows the characterization of the superconducting transition by specific heat measurements. Fig.~\ref{Fig4}(a) shows $\frac {C}{T}$ as a function of $T$, which shows the specific heat jump at the thermodynamic transition. This jump is completely suppressed under 1 T applied magnetic field. The superconducting transition temperature $T_{c}$($x=0.40$) = 1.5 K is shown in the inset of Fig.~\ref{Fig4}(b), as extracted by the standard equal area construction method. The ratio $\frac{\Delta C}{\gamma T_{c}}$ can be used to measure the strength of the electron-phonon coupling.~\cite{padamsee1973} The low temperature normal state specific heat can be well fitted with $\frac{C}{T} = \gamma + \beta T^{2}$, where $\gamma T$ represents the electronic contribution and $\beta T^{3}$ represents the lattice contribution to the specific heat. The solid line in Fig.~\ref{Fig4}(b) shows the fitting; the electronic specific heat coefficient $\gamma (x=0.40) = 4.15 \frac{mJ}{mol~K^{2}}$ and phonon/lattice contribution $\beta (x=0.40) = 0.59 \frac{mJ}{mol~K^{4}}$ are extracted from the fitting. 

The value of $\gamma$ obtained is relatively small, but is slightly higher than that of the cubic tungsten bronze superconductor at approximately the same degree of electron doping.~\cite{daigo2011,zumsteg1976} The low value of $\gamma$ is an indication of a low density of states near the Fermi level, which is a characteristic of the superconducting tungsten bronze family (see Table.~\ref{tab:2}). The specific heat jumps $\frac{\Delta C}{T_{c}}$ for the sample is about 5 mJ mol$^{-1}$ K$^{2}$, setting the values of $\frac{\Delta C}{\gamma~T_{c}}$ to 1.03 for $x=0.40$. This is smaller than the theoretical value of 1.43 for a conventional BCS superconductor. The results suggest that TTB-I K$_{x}$WO$_{3}$ is a weakly electron$-$phonon coupled superconductor.

\begin{figure} [t]
  %\advance\leftskip-4cm
  \centerline{\includegraphics[width=0.4\textwidth]{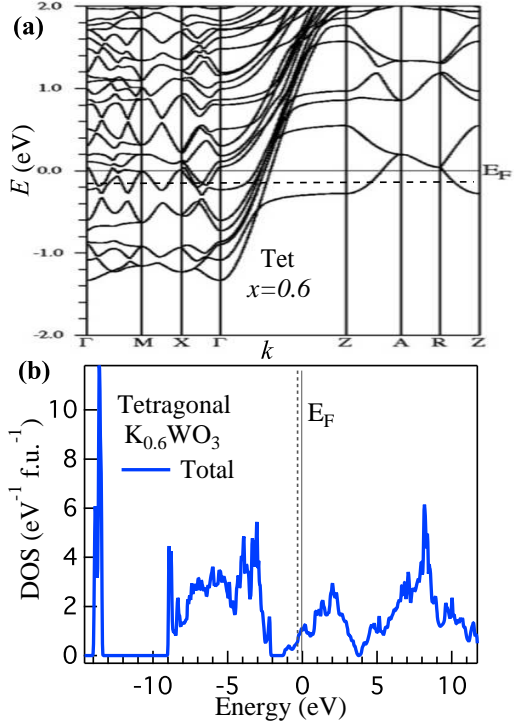}}
  \centering 
  \setlength{\abovecaptionskip}{20pt}
  \caption
    {
      Analysis of the electronic density of states (DOS) for the potassium doped tetragonal tungsten bronze. (a) The electronic structure of the K$_{0.6}$WO$_{3}$ tetragonal tungsten bronze. (b) The total DOS as a function of energy. The dotted line represents the Fermi level corresponding to a $x=0.4$ doping level.
    }
  \label{Fig6}
\end{figure}

\begin{figure*} [htbp]
  %\advance\leftskip-8cm
  \centerline{\includegraphics[width=0.8\textwidth]{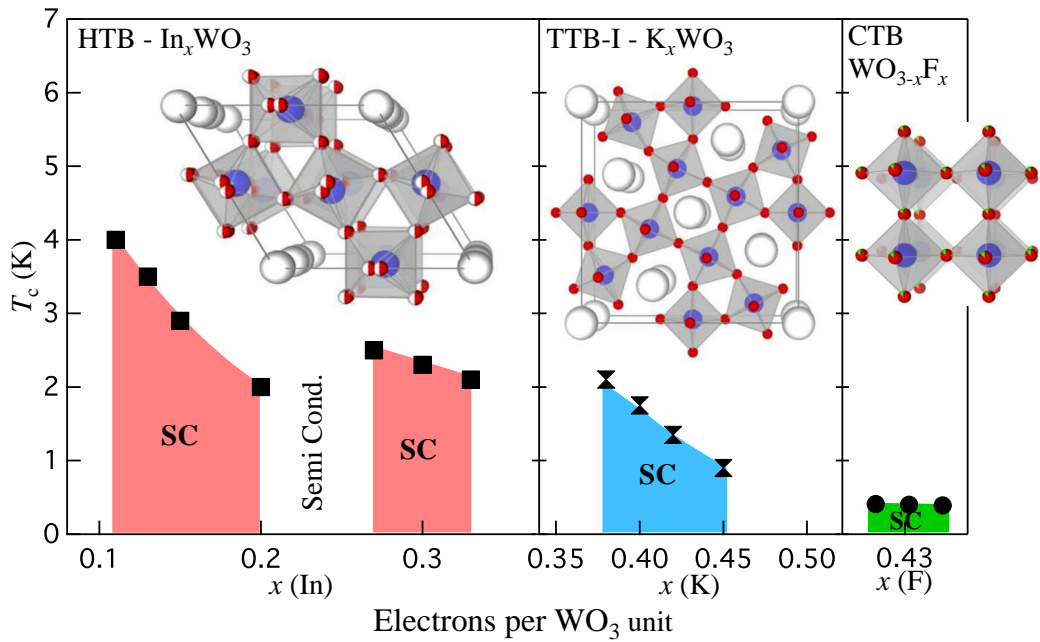}}
  \centering
  \setlength{\unitlength}{\textwidth}
  \setlength{\abovecaptionskip}{10pt}
  \caption
    {
      The superconducting phase diagram for the tungsten bronze family. $T_c$ vs. electron count for the In-doped HTB hexagonal tungsten bronze superconductor, K-doped TTB-I tetragonal tungsten bronze superconductor, and for the F-doped CTB cubic tungsten bronze superconductor, respectively from left to right. The insets show the 3D crystal structures projected on $\textit{ab}$-plane. The phase diagrams for the HTB and CTB are derived from data in ref.14 and ref.15 respectively.    
    }
  \label{Fig5}
\end{figure*}

In a simple Debye model for the phonon contribution to the specific heat, the $\beta$ coefficient is related to the Debye temperature $\Theta _{D}$ through $\beta = N\frac{12}{5}\pi ^{4}R\Theta _{D}^{-3}$, where $R = 8.314~J~mol^{-1}~K^{-1}$. The calculated Debye temperature for K$_{0.40}$WO$_{3}$ is 242.4 K. This value of Debye temperature slightly lower than that of the CTB~\cite{daigo2011,zumsteg1976} and HTB~\cite{bevolo1974} superconductors. 
High values of Debye temperature are often deduced for the tungsten bronze family. This can be explained by including an Einstein term in the heat capacity, due to the rattling of the dopant ion in its cavity. The measurements give only the sum of the framework Debye and dopant ion Einstein contributions.~\cite{bevolo1974} 

\begin{table}[t]
\caption{Superconducting Parameters of the Tungsten Bronze Family.}
  \centering  
  \begin{tabular}{ lc c c c c }
  \hline \hline 
    Parameter & Units & HTB & TTB & CTB \\ 
     &  & (Ref. 24, 22) &  & (Ref. 15)  \\ \hline
      &   & Rb$_{x}$WO$_{3}$ & K$_{0.40}$WO$_{3}$ & WO$_{2.59}$F$_{0.41}$ \\ 
  \hline  
    $T_{c}$ & K & 2.7-4.3 & 1.73 & 0.4  \\
    $\rho _{0}$ & $m\Omega cm$ &  & 0.27 &  \\
    $\frac{dH_{c2}}{dT}\vert _{T=T_{c}}$ & $T~K^{-1}$ & -0.067 to -0.4 & -0.12 & -1.89 \\
    $\mu _{0}H_{c1}(0)$ & T & 53-140 &  &  &   \\
    $\mu _{0}H_{c2}(0)$ & T & 0.11-1.2 & 0.14 & 0.53  \\
    $\mu _{0}H^{Pauli}$ & T & 4.9-7.9 & 3.22 & 0.74 \\
    $\xi (0)$ & \AA & 538-113 & 485 & 250 \\
    $\lambda (0)$ & \AA & 2450-1506 &  &  \\
    $\gamma (0)$ & $\frac{mJ}{mol~K^{2}}$ & 4.2 & 4.15 & 1.59 \\
    $\frac{\Delta C}{\gamma T_{c}}$ & & 1.43 & 1.03 & 1.1  \\
    $\Theta _{D}$ &  & 415 & 242 & 340  \\
    $\lambda _{ep}$ &  &  & 0.52 &   \\  
    $N(E_{F})$ & $\frac{eV}{f.u.}$ &  & 0.27 &  \\ \hline \hline
  \end{tabular}
  \label{tab:2}
\end{table}

An estimation of the strength of the electron-phonon coupling can be derived from the McMillan formula.~\cite{mcmillan1968,poole1999} 
\begin{equation}
\label{equ2} \lambda _{ep} = \frac{1.04 + \mu ^{*} ln\frac{\Theta _{D}}{1.45T_{c}}}{(1-0.62\mu ^{*}) ln\frac{\Theta _{D}}{1.45T_{c}}-1.04}
\end{equation}

McMillan’s model contains the dimensionless electron-phonon coupling constant $\lambda _{ep}$, which, in the Eliashberg theory, is related to the phonon spectrum and the density of states. This parameter $\lambda _{ep}$ represents the attractive interaction, while the second parameter $\mu ^{*}$ accounts for the screened Coulomb repulsion.

Using the Debye temperature $\Theta _{D}$, critical temperature $T_{c}$, and assuming $\mu ^{*} = 0.15$, the electron-phonon coupling constant ($\lambda _{ep}$) obtained for K$_{0.40}$WO$_{3}$ is 0.52, which suggests weak-coupling behavior. This agrees well with $\frac{\Delta C}{\gamma T_{c}} = 1.03$. The value of $\gamma$ extracted from the measured specific heat data corresponds to an electronic density of states at the Fermi energy $N(E_{F})$ of 0.27 states/(eV f.u.) (f.u. = formula unit), as estimated from the relation~\cite{kittel}
\begin{equation}
\label{equ3} \gamma = \pi ^{\frac{3}{2}} k_{B}^{2} N(E_{F}) (1+\lambda _{ep}).
\end{equation}

Assuming that the diamagnetic and vanVleck contributions to the magnetic susceptibility are small, susceptibility at 10 K ($\chi = 1.2 \times 10^{-8} \frac{m^{3}}{mol}$) can be considered as the spin susceptibility and estimate Wilson ratio $R_{W} = \frac{\pi ^{2} k_{B}^{2} \chi _{spin}}{3\mu _{B}^{2}\gamma}$ = 0.6, which is small but close to the free electron value of 1. Also, the coefficient of the quadratic resistivity term can be normalized with the effective mass term from the heat capacity, which gives the Kadowaki-Woods ratio $\frac{A}{\gamma ^{2}}$. This ratio is found to be 0.24~$a_{0}$, where $a_{0}$ = 10$^{-5} \frac{\mu \Omega ~cm}{(mJ / mol~K)^{2}}$. $R_{W}$ and $\frac{A}{\gamma ^{2}}$ both indicate that K$_{0.42}$WO$_{3}$ is a weakly-correlated electron system.~\cite{jacko2009,wilson1975,yamada1975,amar2011}

Fig.~\ref{Fig6} shows the analysis of the electronic density of states for the potassium doped tetragonal tungsten bronze. Fig.~\ref{Fig6}(a) shows the band structure in the vicinity of E$_{F}$, and (b) shows the total DOS, of K$_{0.6}$WO$_{3}$ as a function of energy. 
A similar calculation was done on tetragonal WO$_{3}$ without potassium, suggesting that the rigid band model works well for K doped TTB-I. The dotted lines in Fig.~\ref{Fig6}(a and b) represent the Fermi level for $x=0.4$ doping level, which was found by integrating the DOS.
The band structure calculation clearly shows that near $x=0.35$ the Fermi level transitions through a van Hove singularity (vHs) followed by a flat band between the $\Gamma$ and $Z$ points. This anomalous electronic structure is found close to the change from hexagonal to tetragonal structural phases in the K$_{x}$WO$_{3}$ system. Although it is generally expected that a vHs has a significant influence only on superconductivity in 2D electronic systems~\cite{hirsch1986,dzaloshinskii1987,felser1999,felser1997} the highest $T_{c}$ observed at $x=0.38$ in the doped TTB-I phase appears to be well correlated with the vHs present in the band structure. 
Fig.~\ref{Fig6}(b) shows the potassium doping effect, i.e. of increased electronic concentration, on the insulating WO$_{3}$ system. The Fermi level (E$_{F}$) is shifted upward,~\cite{shein2011,ingham2005} into the bottom of the conduction band. This leads to the metallic behavior of the K$_{x}$WO$_{3}$ TTB-I phase up to the highest possible K doping level $x=0.6$.
 
The superconducting parameters for representative HTB, TTB-I and CTB doped tungsten bronzes are presented in Table.~\ref{tab:2}. Our determined values for the TTB-I phase are employed. The phase diagrams for all the characteristic structure types in the superconducting tungsten bronze family are shown in Fig.~\ref{Fig5}. The diagram shows that the superconducting $T_c$s generally decrease with increasing electron doping of the WO$_3$ framework. It also shows that the superconducting transition temperatures are highest at the lowest doping levels for both the HTB and TTB-I systems; the well-known anomaly in the HTB system near $x$ = 0.25 is displayed.~\cite{josh2013,cadwell1981,2sweedler1965} Finally, it is seen that at comparable doping levels, the perovskite-based CTB is a significantly worse superconductor than the TTB-I bronze phase. This is one of the few systems known where a direct comparison of perovskite-based and non-perovskite-based structure superconductors is possible at the same electron count; in this case the perovskite is the worse superconductor. 

\section{Conclusion}
We have synthesized and characterized the superconducting properties of the K-doped tungsten bronze with the tetragonal-I structure type. A bulk superconducting transition for the TTB-I phase is confirmed through resistivity and heat capacity measurements of K$_{x}$WO$_{3}$. The highest $T_{c}$ observed is 2.1 K. The electronic contribution to the specific heat is relatively small, $\gamma$ = 4.15 $\frac{mJ}{mol~K^{2}}$ and the electron-phonon coupling constant $\lambda _{ep}=0.52$. The electronic density of states calculations indicate that close to the HTB-TTB structural phase boundary, the Fermi Level in the TTB-I phase transitions through a van Hove singularity at $x=0.35$; the highest $T_{c}$ in the tetragonal-I system is at a correspondingly low doping level.   
Finally, with this characterization of the TTB-I superconductor, it is now possible to present a complete superconducting phase diagram for the tungsten bronze family. The highest superconducting $T_{c}$ is observed in general at the lowest electron doping level for the WO$_3$ framework, and also at the lowest doping level for the HTB and TTB-I superconductors individually; it is almost independent of doping for the CTB. The cubic bronze phase is less favored for superconductivity than the tetragonal bronze phase, where the connectivity of the WO$_6$ octahedra is different from what is seen in the perovskite. The data provided suggest that a global theoretical treatment of all the compounds in this family of superconductors would be of interest.

\section{acknowledgments}
This work was supported by the Department of Energy Division of Basic Energy Sciences, grant DE-FG02-98ER45706. NH would like to thank Mazhar~N.~Ali and T.~Klimczuk for useful discussions.

\end{document}